\documentclass[11pt,showpacs,preprintnumbers,amsmath,amssymb]{revtex4}

\usepackage{graphicx}
\usepackage{dcolumn}
\usepackage{bm}

\begin{document}

\preprint{KEK Preprint 2005-56}

\title{Efficient Propagation of the Polarization from Laser Photons to 
Positrons \\through Compton Scattering and Electron-Positron 
Pair Creation}

\author{T. Omori$^{1}$}
\author{M. Fukuda$^{2}$}
\author{T. Hirose$^{3}$}
\author{Y. Kurihara$^{1}$}
\author{R. Kuroda$^{3, 4}$}
\author{M. Nomura$^{2}$}
\author{A. Ohashi$^{5}$}
\author{T. Okugi$^{1}$}
\author{\\K. Sakaue$^{3}$}
\author{T. Saito$^{3}$}
\author{ J. Urakawa$^{1}$}
\author{M. Washio$^{3}$}
\author{I. Yamazaki$^{3}$}
\affiliation{$^{1}$KEK: High Energy Accelerator Research Organization, 
             Tsukuba-shi, Ibaraki 305-0801, Japan\\
             $^{2}$National Institute of Radiological Sciences, Chiba-shi, 
             Chiba 263-8555, Japan\\
             $^{3}$Advanced Research Institute for Science and Engineering, 
             Waseda University, Tokyo 169-8555, Japan\\
             $^{4}$National Institute of Advanced Industrial Science and 
             Technology, Tsukuba-shi, Ibaraki 305-8568, Japan\\
             $^{5}$Department of Physics, Tokyo Metropolitan Univeresity, 
             Hachioji-shi, Tokyo 192-0397, Japan}

\date{February 21, 2006}

\begin{abstract}
We demonstrated for the first time the production of highly polarized 
short-pulse positrons with a finite energy spread in accordance with a 
new scheme that consists of two-quantum processes, such as inverse 
Compton scattering and electron-positron pair creation. 
Using a circularly polarized laser beam of 532 nm scattered off a 
high-quality, 1.28 GeV electron beam, we obtained polarized positrons 
with an intensity of $2 \times 10^4$ $e^+$/bunch. The magnitude of positron 
polarization was determined to be $73\pm15(sta)\pm19(sys)\%$ 
by means of a newly designed positron polarimeter.
\end{abstract}

\maketitle

{\it Introduction.}---
A positron($e^+$), the antiparticle of an electron($e^-$), is created 
from the pair creation by high-energy photons and can also be obtained 
from beta decays of specific radioisotopes. Concerning polarization, 
positrons emitted from beta decays are longitudinally polarized but are 
subject to a large energy spread, a wide angular distribution, 
low intensity, etc, such that those positrons are far from a practical 
usage as an $e^+$ beam. It is widely believed that one of the 
next-generation accelerators at the energy frontier will be 
an electron-positron linear collider, the International Linear 
Collider (ILC), where the polarized $e^+$ beams should play 
significant roles in studying the standard model as well as 
discovering phenomena beyond the standard model.

\begin{figure}
\scalebox{.50}{\includegraphics{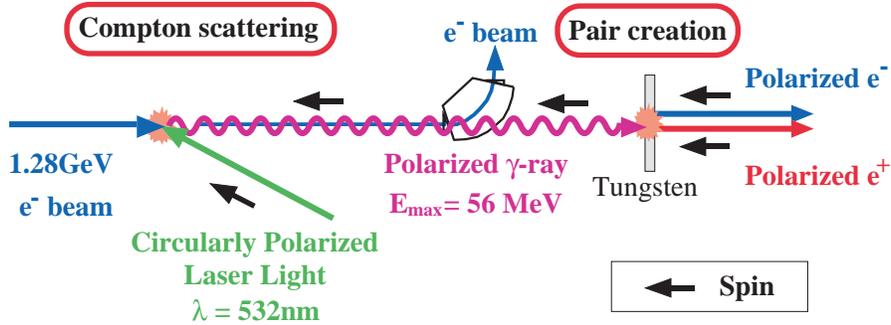}}
\caption{\label{fig:fig1} A fundamental scheme of polarized $ e^+$ 
production. Right-handed polarized laser photons are backscattered 
off relativistic electrons resulting in production of left-handed 
polarized $\gamma$-rays in the forward direction (in the high energy 
part of the spectrum). Pair creation of the $\gamma$-rays through 
a tungsten plate generates left-handed positrons in the high energy part.}
\end{figure}

\begin{figure}
\scalebox{.45}{\includegraphics{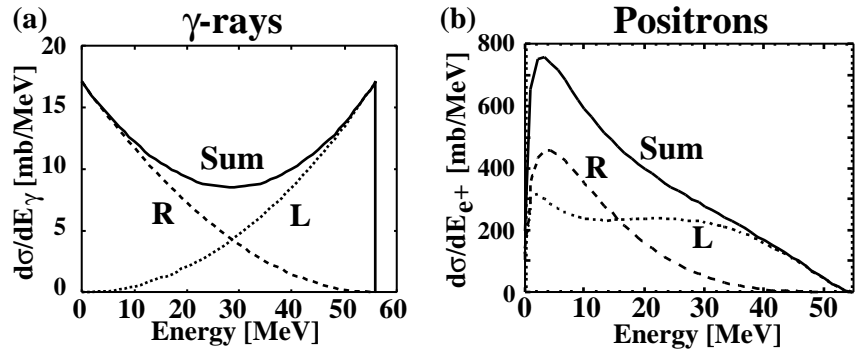}}
\caption{\label{fig:2} (a) Differential cross section of the Compton 
scattering for right-handed polarized laser photons with wavelength of 
532 nm backscattered off 1.28 GeV electrons as a function of 
the $\gamma$-ray energy. The dashed and dotted curves 
correspond to the helicities of $+1$ and $-1$ for the gamma-rays, 
respectively.  
(b) Differential cross section of the positron creation (electron-positron 
pair creation) in the thin tungsten target calculated using the 
$\gamma$-ray distribution given in fig.2(a).  The dashed 
and dotted curves correspond to the helicities of $+1$ and 
$-1$ for the positrons.}
\end{figure}

In 1996 we proposed a novel method\cite{OkugiJJAP1996} for creating 
highly polarized $e^+$ beams with a finite energy spread through 
electron-positron pair creation from longitudinally polarized 
$\gamma$-rays that are generated from the inverse Compton scattering 
of circularly polarized laser-photons. 
Fig.~\ref{fig:fig1} is a schematic diagram of the proposed method where 
the helicity of a laser photon is +1 (right-handed, R), for example. 
Fig.~\ref{fig:2}(a) shows the energy dependence of the cross section 
of the inverse Compton scattering in which a right-handed polarized 
laser light of wavelength 532 nm scatters off a relativistic electron 
beam with an energy of 1.28 GeV. Note that the left-handed $\gamma$-rays 
dominate in the energy close to the maximum energy of 56 MeV. 
When these left-handed $\gamma$-rays hit on a thin metallic target, 
highly polarized positrons are generated in the high-energy part of 
the spectrum, as can be seen in Fig.~\ref{fig:2}(b). We therefore are able to 
accomplish the efficient propagation of the polarization from the 
laser photon to the positron via two quantum processes.

We have pursued proof-of-principle studies with the help of extensive 
Monte-Carlo simulations\cite{HiroseNIM2000,SakaiPRSTAB2003}.  
One of the important steps is to establish the polarimetry for 
short-bunch $\gamma$-rays and positrons with a time duration of a 
few tens of picoseconds (ps). Usually, the $\gamma$-ray polarization 
can be determined using spin-dependent Compton scatterings of 
$\gamma$-rays on electrons in 3d orbits in magnetized iron. 
Actually, in this experiment, the magnetized iron is a 100 mm long 
iron pole of an electro magnet (called the analyzer magnet hereafter). 
However, the electron-positron pair-creation in the analyzer magnet 
causes considerably large backgrounds that become a serious obstacle 
to cleanly extracting the Compton process. Thus, the Compton process 
has to be selected by means of a coincidence method in which both 
the electrons and $\gamma$ photons emerging at the Compton scattering 
should be simultaneously identified. Unfortunately, such a 
short-bunch width of the $\gamma$-ray beam, only a few tens of 
picoseconds, prevents us from applying the coincidence method.

We adopted a ``transmission method'' in which only the intensity of 
the transmitted $\gamma$-rays, $N_P$ for the parallel cases and $N_A$ 
for the anti-parallel cases, was measured at a position downstream 
of the magnet\cite{CulliganNature1957, GoldhaberPR1958, MacqPR1958}. 
Here the parallel (anti-parallel) 
case means that the spin of the $\gamma$-rays and that of the electrons 
in the analyzer magnet are parallel (anti-parallel) to each other. 
Hence the asymmetry is defined as $A = (N_P - N_A)/ (N_P + N_A)$. 
The asymmetry is also the product of the polarization $P$, and $A_{MC}$ , 
the analyzing power estimated from the Monte Carlo 
simulation ($A = PA_{MC}$). Because the transmitted $\gamma$-rays 
are collimated to a narrow forward-cone, whereas the Compton 
$\gamma$-rays and the pair-created electrons and positrons as 
sources of background are emitted over a wide angular region, 
influence of such backgrounds is small enough that the transmission 
method can be applied to polarization measurements of 
$\gamma$-rays with any time structure.

{\it Polarized positron production.}---
A series of experiments has been conducted at the KEK-ATF, 
(Accelerator Test Facility)\cite{KuboPRL2002}, consisting of an 
S-band linac and a damping ring, which provides high-quality 
1.28 GeV $e^-$ beams, with a typical beam intensity of 
$1.8\times10^{10}$ $e^-$/bunch, a repetition rate of 3.12 Hz 
and an extremely small emittance. A mode lock Nd:YAG laser produces 
second-harmonic laser light with a wavelength of 532 nm and an average 
energy of 400 mJ/pulse. The laser light is converted into 
circularly polarized photons while passing through a quarter wave 
plate. The spot sizes in r.m.s. at the collision point are around 
60 $\mu$m (horizontal) and 40 $\mu$m (vertical) for the $e^-$ beam 
and 50 $\mu$m for the round laser beam. The Compton back scattering 
of the polarized laser beam with the $e^-$ beam generated polarized 
$\gamma$-rays with a maximum energy of 56 MeV and a bunch length 
of 31 ps, being equal to the bunch length of the $e^-$ beam.

\begin{figure}
\scalebox{.45}{\includegraphics{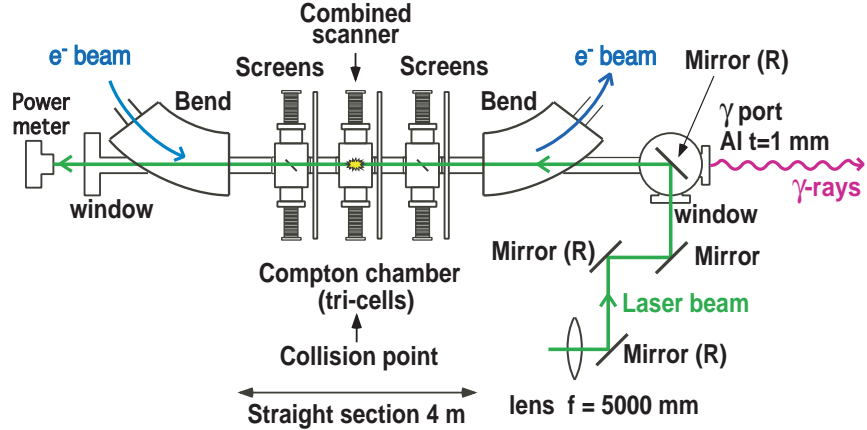}}
\caption{\label{fig:3} Experimental setup of the polarized 
$\gamma$-ray production system including the Compton chamber with 
three cells, the laser optics with three remotely controllable 
mirrors, and electron beam line. Laser-electron collisions take place 
at the central cell of the Compton chamber.}
\end{figure}

Fig.~\ref{fig:3} shows an apparatus, called a Compton chamber, consisting of 
three cells together with the electron and laser beam lines. In order 
to achieve precise diagnoses for both beams, we installed screen monitors, 
a wire scanner and a knife-edge scanner in the Compton 
chamber\cite{FukudaPRL2003}. 
To observe the transverse position and the angle of both beams, 
the screen monitors were mounted in the central cell located at the 
collision point and in the side cells placed 265 mm away from the 
collision point. The wire scanner and the knife edge scanner were 
set in the central cell. The wire scanner allows us to measure the 
$e^-$ beam sizes at the collision point; and the knife edge scanner 
determines the laser profile. The laser beam is transported to the 
collision point over a distance of about 10 m using six mirrors 
coated with a multilayer dielectric. The three mirrors can be remotely 
controlled to allow adjustment the position and angle of the laser 
beam at the collision point. Technical details on the production of 
polarized $\gamma$-rays and the $\gamma$-ray polarimetry may be 
found in Ref.~\onlinecite{FukudaPRL2003}.

\begin{figure}
\scalebox{.45}{\includegraphics{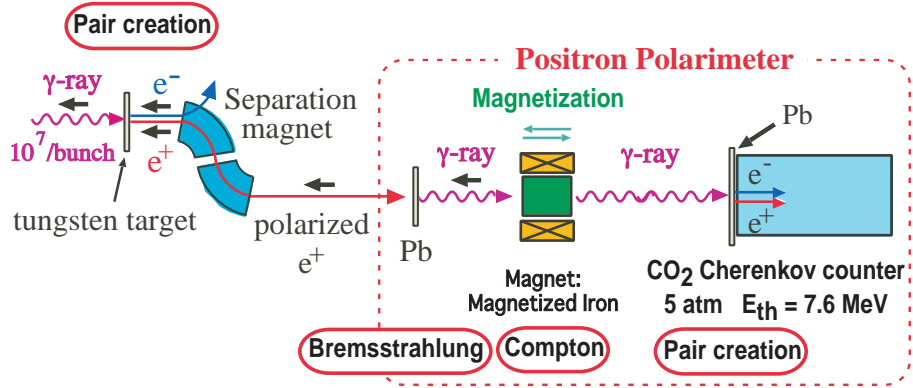}}
\caption{\label{fig:4} Schematic drawing for the $e^+$ production 
apparatus and the polarimeter. Polarized $\gamma$-rays are incident 
on a 1mm thick tungsten plate resulting in electron-positron pair 
creation. The separation magnet installed after the tungsten plate 
efficiently extracts high energy positrons, which generates 
circularly polarized $\gamma$-rays through bremsstrahlung in the lead. 
The transmission of the $\gamma$-rays can be measured with the 
CO$_2$-gas Cherenkov counter.}
\end{figure}

To maximize the collision probability of the laser photons and 
electrons, we optimized the laser optics as follows. The original laser 
beam was expanded to a spot size of 4.6 mm in r.m.s. and transported to 
the last lens with a relatively long focal-length, 5000 mm, resulting 
in a long collision distance of about 7 cm. Furthermore, to enable 
head-on collisions, the last mirror of 3 mm thickness was placed on 
the axis of the $\gamma$-rays, as shown in Fig.~\ref{fig:3}. Note 
that loss of generated $\gamma$-rays was negligible while passing through 
the mirror. The average number of polarized $\gamma$-rays generated in 
the inverse Compton process was $2 \times 10^7$/bunch. 
These $\gamma$-rays were 
incident on a 1 mm thick tungsten plate for electron-positron pair 
creation. A separation magnet consisting of a pair of dipole magnets 
was installed after the tungsten plate, as shown in Fig.~\ref{fig:4}, 
to efficiently extract high-energy positrons that should have a high 
degree of polarization. We thus obtained $2 \times 10^4$ $e^+$/bunch. 
The numbers of $\gamma$-rays and positrons were measured by 
an air Cherenkov counter and Si PIN photodiodes, respectively. 
While we did not measure the momentum of the 
positrons, according to simulation, the average energy of positrons 
is 36 MeV with r.m.s. width of 8 MeV.

{\it Positron polarimeter.}---
Extending the transmission method employed in the $\gamma$-ray 
polarimetry\cite{FukudaPRL2003}, we have designed a $e^+$ polarimeter 
whose schematic view is depicted in Fig.~\ref{fig:4}. 
The separated positrons hit a 6 mm thick lead converter, in which 
longitudinally polarized $\gamma$-rays are 
again generated via the bremsstrahlung process from the positrons. 
The polarization of these $\gamma$-rays can be measured by means 
of $\gamma$-ray polarimetry, the transmission method, which allows 
determination of the $e^+$ polarization. The transmission, T, is 
defined as the number of transmitted $\gamma$-rays normalized to 
the number of $\gamma$-rays generated from inverse Compton scattering. 
In actual data treatment, an asymmetry is determined by flipping 
the polarity of the analyzer magnet. For a fixed state of the laser 
polarization, this corresponds a fixed state of the positron 
polarization, we measured the transmissions $T_+$ and $T_-$ for 
the magnetization direction being upstream and downstream of the 
positron flight direction, respectively.  Thus the asymmetry, A, 
is given as $A = ( T_+  - T_- ) / ( T_+ + T_- )$.

For checking the consistency of the whole system, measurements were 
carried out by reversing the laser polarization. Suppresion of 
backgrounds from $e^-$ beam halo interacting with the beam pipe and 
material surrounding it is critical to measurement as the asymmetry 
is very small ($\sim 1 \%$). Lead blocks were placed at appropriate 
locations along the beam line, and five-fold collimators were set 
between the collision point and the tungsten target to suppress the 
backgrounds. The $e^+$ polarimeter was shielded with 20 cm thick 
iron walls in the front and 10 cm at the side. The photomultiplier 
of the CO$_2$ Cherenkov counter was heavily shielded with lead 
blocks and was placed on the floor to escape from background 
produced by $e^-$ beam which was running at 1.2 m high. 
Also the gate width of the ADC to read the photomultiplier was 
minimized to 40 ns.

In order to maximize the spatial overlap between the $e^-$ and 
laser beams, we precisely adjusted the position and angle of the 
laser beam using the three remotely controllable mirrors shown in 
Fig.~\ref{fig:3}. The collision timing was set such that the 
$\gamma$-ray signal after the Compton scatterings became largest 
by controlling the laser timing.

Taking into account all the elements of the polarimeter shown 
in Fig.~\ref{fig:4}, we performed various simulations to estimate 
the expected asymmetry with the help of following 
simulation-codes: CAIN\cite{YokoyaCAIN} for Compton scattering of 
laser beams on $e^-$ beams; GEANT3\cite{GEANT3} for electromagnetic 
interactions in a material; GRACE\cite{GRACE} particularly for 
spin-dependent electron-gamma interactions in the analyzer 
magnet ; and POISSON\cite{POISSON} for iron magnetization.

\begin{figure}
\scalebox{.60}{\includegraphics{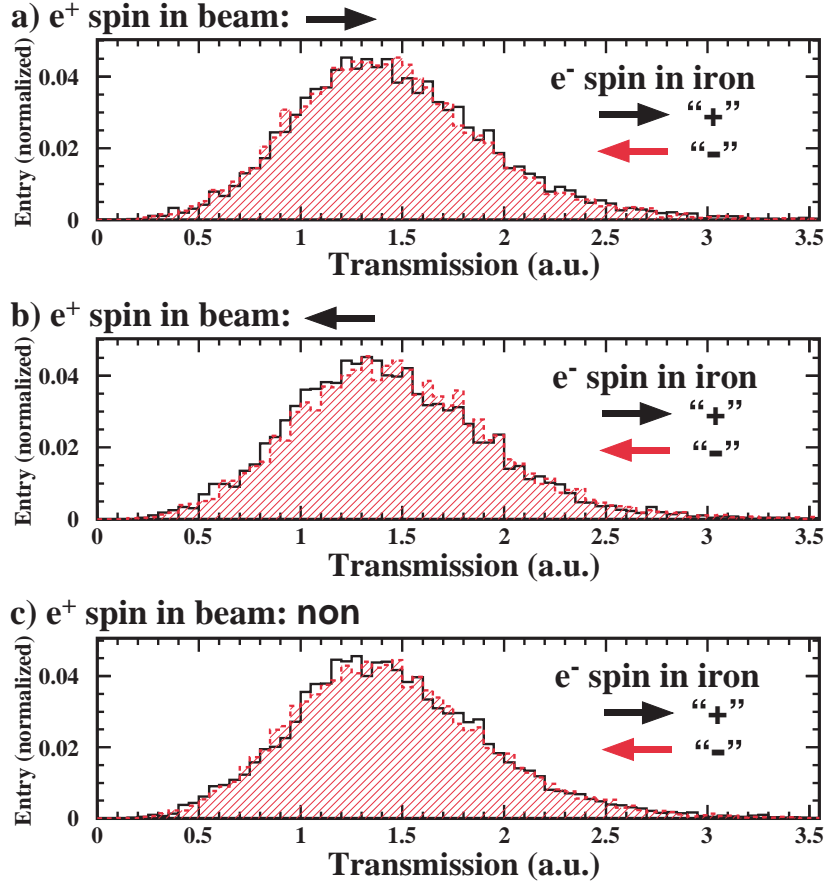}}
\caption{\label{fig:5} Measured asymmetries of polarized positrons 
for the different states of laser polarization (positron polarization).  
(a) Left-handed laser polarization (right-handed positron polarization).  
(b) Right-handed laser polarization (left-handed positron polarization). 
(c) Linear laser polarization (positrons are not logitudinaly polarized). 
In each figure, black-line histogram and red-hatched histograms show 
distributions corresponding to magnetization directions being 
upstream (``+'') and downstream (``$-$'') of the positron 
flight-direction}
\end{figure}

\begin{figure}
\scalebox{.320}{\includegraphics{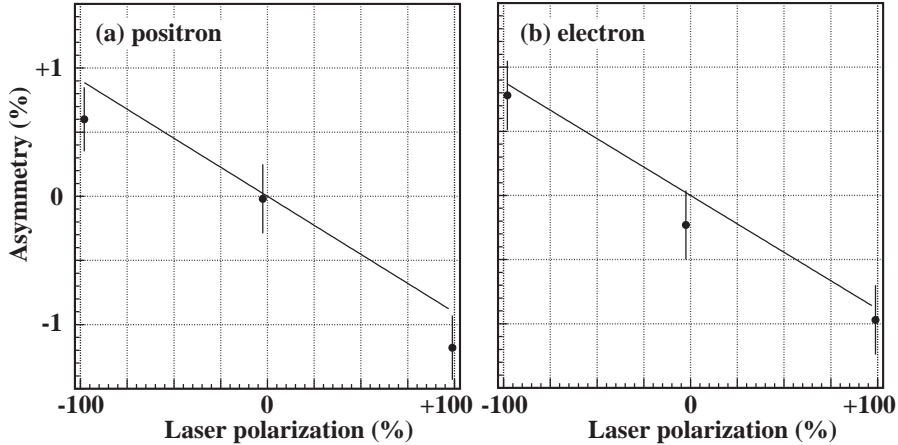}}
\caption{\label{fig:6} Measured asymmetry as a function of the circular 
polarization of a laser beam. Straight lines are result of linear fitting 
with a constraint of passing the origin (0,0); (a) result of the 
positron measurement, (b) result of the electron measurement. 
Note: When the direction of photon spin and the direction of it's 
momentum is paralell, we call it right-handed. When a laser beam is 
$100\%$ right-handed polarized, we call it $+100\%$.}
\end{figure}

{\it Results and conclusions.}---
Fig.\ref{fig:5}(a) shows result of asymmetry measurement when laser 
polarization set to be left-handed. The transmission distributions for 
the ``+'' and ``$-$'' magnetization directions are shown by lack line 
and red-hatched histograms, respectively. 
The significant deviation between the two 
distributions can be observed, leading to an asymmetry of $0.60\pm0.25\%$. 
We conducted the same measurement by reversing the laser polarization 
and obtained an asymmetry of $-$$1.18\pm0.27\%$ as given in 
Fig.\ref{fig:5}(b).  In order to check the consistency of the 
measurement, we also conducted the asymmetry measurement for linear 
laser polarization, which should correspond to zero longitudinal 
polarization of positrons: The results of this run are shown in 
Fig.\ref{fig:5}(c), from which we obtained an asymmetry of 
$-0.02\pm0.25\%$, which is consistent with zero polarization of 
positrons. Here, all errors noted above are statistical ones. 
The results of asymmetry measurement corresponding to three laser 
polarization shown in Fig.\ref{fig:5}(a)(b)(c) are plotted in 
Fig.\ref{fig:6}(a).  
As shown in Fig. 6(a), we fit a straight line to the data. In the fitting
the line is constrained to pass through the origin (0,0).
Hence the absolute value of the slope is obtained to be 
$0.90\pm0.18\%$ with a reduced  $\chi^2$ of 1.3. 
Here the error represents the fitting error and therefore it is 
essentially statistical. This slope corresponds to the positron 
asymmetry with $100\%$ laser polarization. From this asymmetry, 
the magnitude of the positron polarization was calculated as 
$73\pm15\pm19\%$, where the first error is statistical one 
(the fitting error) and the second error is systematic one which comes 
from the uncertainty in a Monte-Carlo simulation. The measured value 
of the polarization is consistent with the value of $77\%$ estimated 
by a Monte-Carlo simulation. 

For further confirmation, we also produced polarized electrons 
by reverting the polarity of the separation magnet (see Fig.~\ref{fig:4}) 
and performed polarization measurements.
The measured 
asymmetries were $0.78\pm0.27\%$ and $-0.97\pm0.27\%$ for left-handed 
and right-handed laser polarization, respectively. And for the linear 
laser polarization, the asymmetry was measured to be $-0.23\pm0.27\%$.  
The results of electron measurements are plotted in Fig.\ref{fig:6}(b) 
with result of linear fitting. The asymmetry calculated from the 
fitting is $0.89\pm0.19\%$, with reduced $\chi^2$ of 0.55.  
The data of positron polarization and electron polarization are in 
agreement with each other.

The present results have demonstrated for the first time the viability 
of our proposed scheme, which is based on two quantum processes, 
namely inverse Compton scattering of circularly polarized laser 
light on high-quality electron beams and electron-positron pair 
creation. It has been verified that efficient propagation of the 
polarization to the positrons by imposing specific kinematical 
conditions, specially the selection of relatively high energy 
positrons. Using a circularly polarized laser beam of 532 nm scattered 
off a high-quality 1.28 GeV electron beam, we obtained polarized 
gamma-rays of $2 \times 10^7$/bunch and polarized positrons of 
$2 \times 10^4$/bunch. 
We have established positron polarimetry based on the so-called 
transmission method and have obtained a transmission asymmetry 
of $0.90\pm0.18\%$ leading to a positron polarization of 
$73\pm15(sta)\pm19(sys)\%$. This value is in good agreement with 
the value of $77\%$ derived from Monte-Carlo simulations in which 
basic QED processes and beam parameters are involved. Various 
techniques developed in this study for finding the optimal condition 
for laser-electron beam collisions will be of great use to the 
development of extremely precise beam diagnoses required by future 
linear colliders or third-generation electron accelerators. 
The present results encourage the realization of polarized 
positron beam\cite{OmoriNIM2003} for the future linear collider ILC.

We would like to thank all members of the ATF group for the operation 
of ATF-accelerator.  We also acknowledge Mr.~K.~Sugiyama, Dr.~K.~Dobashi, 
Dr.~I. Sakai, Dr.~T.~Muto, Ms.~A.~Higurashi, Mr.~T.~Iimura, Mr.~T.~Aoki 
for their important contributions in past in developing the experimental 
apparatus, the electron and laser beam optics, and the data analysis. 
This research was partially supported by a Grant-in-Aid for Scientific 
Research (B)11554010, (A)11694092, (C)10640294, by a research 
program of U.S.-Japan Cooperation in the Field of High-Energy Physics, 
and by Advanced Compact Accelerator Project of National Institute of 
Radiological Sciences.


\end{document}